\documentclass[reprint,
superscriptaddress,
amsmath,amssymb,aps,showkeys,showpacs,
twoside,final,secnumarabic,
nofootinbib]{revtex4-2}

\usepackage[paperwidth=205mm,paperheight=290mm,top=17mm,bottom=25mm,
inner=17mm,outer=17mm,
twoside]{geometry}

\usepackage{cmap} 
\usepackage[T1,T2A]{fontenc}
\usepackage[utf8]{inputenc}
\usepackage[russian,english]{babel}
\usepackage{color}
\usepackage{graphicx}
\usepackage{dcolumn}
\usepackage{bm} 
\usepackage[unicode=true,colorlinks=true,linkcolor=magenta, urlcolor=blue, citecolor = blue,breaklinks]{hyperref}
\usepackage{multirow}
\usepackage{url}
\usepackage{breakurl}

\DeclareGraphicsExtensions{.eps}

\newcount\issue
\newcount\Vol
\newcount\numb
\usepackage{fancyhdr} 
\def\Vol{\textbf{78}}
\def\numb{x}
\begin{document}

\title{JOURNAL SECTION OR CONFERENCE SECTION\\[20pt]
Updates from the COSINE experiment} 

\def\addressa{Department of Physics, Chung-Ang University, Seoul, 06974, Republic of Korea}

\author{\firstname{Chang Hyon}~\surname{Ha} on behalf of the COSINE-100 collaboration}
\email[E-mail: ]{chha@cau.ac.kr }
\affiliation{\addressa}

\received{xx.xx.2025}
\revised{xx.xx.2025}
\accepted{xx.xx.2025}

\begin{abstract}
  The COSINE project aims to independently test the DAMA experiment’s long-standing claim of detecting dark matter interactions
  using the same NaI(Tl) scintillating crystal technology. In its first phase, COSINE-100 collected over six years of data at the Yangyang Underground Laboratory
  and achieved a three-sigma exclusion of the DAMA modulation signal under standard dark matter assumptions.
  The experiment is now being upgraded to the COSINE-100 Upgrade, deploying higher–light-yield NaI(Tl) detectors
  at the new, deeper Yemilab facility to further enhance sensitivity. Here, I will discuss the model-dependent WIMP searches performed
  with COSINE-100 and provide an update on the ongoing upgrade effort, highlighting its future prospects for crystal-based dark matter direct detection in Korea.

\end{abstract}

\pacs{Suggested PACS}\par
\keywords{Dark Matter, WIMP, NaI(Tl) crystal \\[5pt]}

\maketitle
\thispagestyle{fancy}


\section{Introduction} \label{intro}
\begin{figure*}[!htb]
  \begin{center}
      \includegraphics[width=0.9\textwidth]{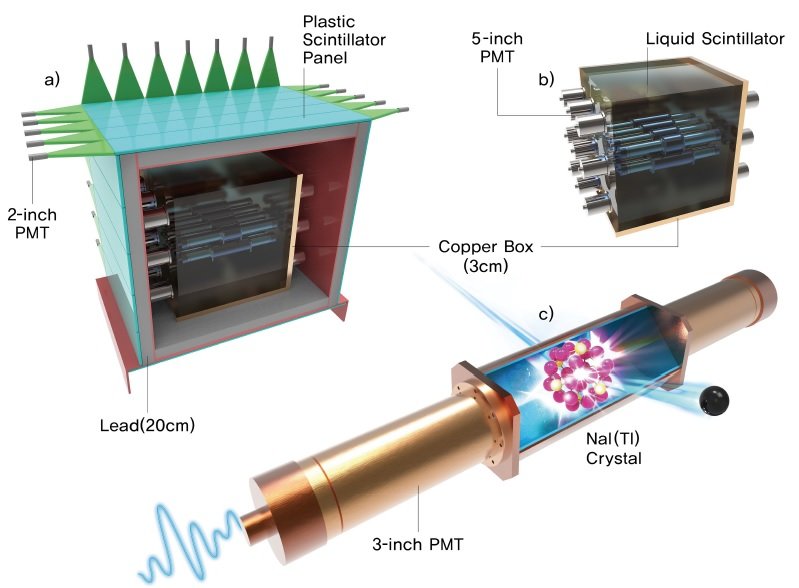}
  \end{center}
  \caption{COSINE shielding overview.
    From outside inward,
    muon panels (3\,cm thick, light blue),
    a lead brick castle (20\,cm thick on all sides, grey),
    a copper box (3\,cm thick),
    an acrylic box (1\,cm thick) and
    eight encapsulated crystal detectors immersed in the liquid scintillator ($>$40\,cm from crystal assembly to wall on all sides)
    are shown.
    Also indicated are the locations of the calibration holes, the size of the PMTs, and labeling scheme for the different sides.
  }
  \label{ref:shield}
\end{figure*}

Although dark matter appears to permeate the Universe, its fundamental properties remain largely unknown. A wide range of astronomical observations—including stellar and galactic rotation curves, anisotropies in the cosmic microwave background, and gravitational lensing—indicate that roughly 26 \% of the Universe is composed of dark matter~\cite{Huterer2010,Clowe:2006eq,Ade:2015xua}. Among the leading particle candidates are Weakly Interacting Massive Particles (WIMPs)~\cite{PhysRevLett.39.165,Jungman:1995df}. Theoretical studies suggest that the rare scatterings of WIMPs in the Milky Way halo with nuclei in terrestrial detectors could be observable with ultra–low-background instrumentation operated deep underground~\cite{Goodman:1984dc}.

A characteristic prediction of such WIMP–nucleon interactions is an annual modulation of the nuclear-recoil rate, arising from the Earth’s orbital motion through the Galactic halo~\cite{PhysRevD.33.3495,Freese:2012xd}. Since 1995, the DAMA/NaI and DAMA/LIBRA experiments (collectively referred to as DAMA) have searched for this signature using an array of radiopure NaI(Tl) crystals~\cite{Bernabei:2008yh}. Across multiple phases, DAMA has consistently reported a statistically significant annual modulation with a phase compatible with expectations from the Earth’s motion relative to the Galactic rest frame~\cite{bernabei00,Bernabei:2013xsa}. Their latest result, based on a total exposure of 2.86\,ton$\cdot$yr using detectors achieving background levels of 1\,count/day/kg/keV\,\footnote{keV denotes electron-equivalent energy.} reports a 13.7\,$\sigma$ modulation in single-hit events between 2–6\,keV, with an amplitude of 0.00996$\pm$0.00074\,count/day/kg/keV when the period and phase are fixed to one year and 152.5\,days, respectively~\cite{Bernabei:2025zsa}.

Despite the long-standing signal, the interpretation of the DAMA modulation as WIMP–nucleon scattering remains highly contentious~\cite{Freese:2012xd,bernabei00,Nygren:2011,Savage:2006qr,Kopp:2009et,Ralston:2010bd}. Under the standard Galactic halo model~\cite{Freese:2012xd}, the cross sections required to explain DAMA are excluded by null results from several leading time-integrated dark-matter searches, including LUX-ZEPLIN~\cite{LZ:2024zvo}, PandaX~\cite{PandaX-4T:2021bab}, XENON~\cite{XENON:2024wpa}, SuperCDMS~\cite{PhysRevLett.112.241302}, and KIMS~\cite{sckim12}. In addition, annual-modulation searches by XMASS~\cite{Abe:2015eos} and XENON~\cite{PhysRevLett.118.101101}, interpreted under leptophilic interaction models, also contradict the DAMA observation.

To clarify this tension, independent experiments using the same target material and similar detector methodology have been developed to provide direct, model-independent tests of the DAMA claim. The two leading efforts—ANAIS~\cite{amare14A} and COSINE-100~\cite{Adhikari:2017esn}—have performed high-precision searches for annual modulation in NaI(Tl) detectors, aiming to confirm or refute the DAMA observation. Both collaborations have released full data analyses~\cite{Amare:2025dfq,COSINE-100:2024nfa} as well as a recent combined result~\cite{ANAIS-112:2025fne}. These measurements collectively show modulation amplitudes consistent with zero at significance levels exceeding 3\,$\sigma$, thereby strongly disfavoring the DAMA result within a model-independent framework.

Given the strong constraints from these modulation analyses, attention has increasingly shifted toward model-dependent interpretations, where additional particle or interaction hypotheses may reconcile DAMA with other experiments or reveal new regions of parameter space accessible to NaI(Tl)-based detectors. In this context, COSINE-100 plays a central role: its low-background NaI(Tl) array offers competitive sensitivity to a broad class of WIMP interaction models beyond the simple elastic, spin-independent scenario.

In this paper, we review the model-dependent dark-matter searches performed with COSINE-100 data. Our goal is to assess COSINE-100’s current standing in the broader landscape of WIMP searches and to outline the future directions and discovery potential of NaI(Tl) dark-matter experiments.

\begin{figure*}[!htb]
  \begin{center}
      \includegraphics[width=0.90\textwidth]{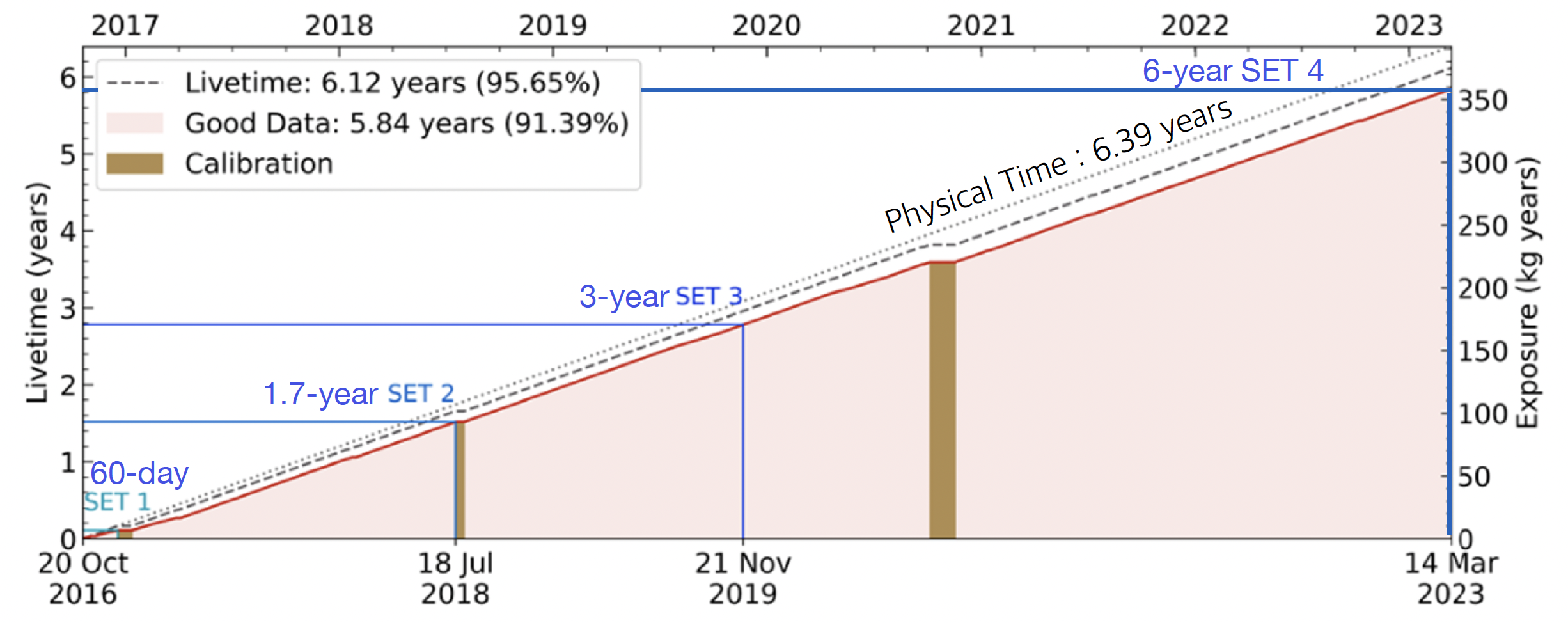}
  \end{center}
  \caption{
    COSINE-100 exposure summary. Over the total run period of 6.39 years, the detector operated for 6.12 years, with 5.84 years of data passing all WIMP-search quality requirements. Periodic calibrations account for the remaining downtime. The data set is analyzed in four subsets (SET1–SET4), as indicated.
  }
  \label{ref:expo}
\end{figure*}
\section{ The COSINE-100 Experiment}
The COSINE experiment is a joint effort of the KIMS and DM-Ice collaborations to test the DAMA claim using the same NaI(Tl) target material. COSINE-100 consists of an array of eight low-background NaI(Tl) crystals with a total mass of 106 kg. Detector construction and assembly were completed in early 2016 at the Yangyang Underground Laboratory (Y2L), and stable physics data-taking began in October 2016. The detector operated continuously until March 2023, delivering more than six years of high-quality data for dark-matter searches.

Y2L is located within the A5 tunnel of the Yangyang pumped-storage hydroelectric power facility, approximately 150 km east of Seoul. The laboratory is accessed via a 2 km horizontal tunnel and is shielded by at least 700 m of rock overburden. The adjacent A6 tunnel previously hosted the KIMS experiment, which operated a CsI(Tl) array for more than 15 years.

COSINE-100 is housed inside a four-layer shielding system designed to suppress external backgrounds as effectively as possible, as shown in Fig.~\ref{ref:shield}. A steel frame supports an inner volume measuring approximately $\rm 300 cm \times 220 cm \times 270 cm$, providing full $4\pi$ coverage. From the outside inward, the detector is enclosed by 3-cm-thick plastic-scintillator panels that serve as a cosmic-ray muon veto, followed by a 20-cm-thick lead-brick castle that attenuates ambient gamma radiation. Inside the lead shield is a 3-cm copper box that suppresses radioactivity originating from the lead and the environment.

At the innermost boundary, the NaI(Tl) crystals are immersed in a 2200-L liquid-scintillator (LS) volume based on Linear Alkyl Benzene (LAB). This LS veto serves as a key feature of the COSINE-100 design: it actively tags external gamma rays, multiple-scattering events, and other backgrounds that would otherwise mimic dark-matter–induced signals in the crystals. Eighteen 5-inch photomultiplier tubes (PMTs) mounted on the sides of the acrylic LS container collect scintillation light from this veto region. The arrangement ensures that the crystals are separated from any surrounding boundary by more than 40 cm of scintillating material, significantly enhancing background rejection capability.

The eight encapsulated NaI(Tl) crystals are mounted on a central support table within the LS veto and optically coupled to their own 3-inch PMTs. A detailed description of the COSINE-100 detector design, materials, and assembly procedures can be found in the dedicated instrumentation paper~\cite{Adhikari:2017esn}.

\section{Analysis and Results}

COSINE-100 collected a total exposure of 360\,kg$\cdot$yr between 20 October 2016 and 14 March 2023. Over this 6.39-year period, the detector operated with excellent stability: more than 91\% of the accumulated data are classified as good-quality runs after excluding calibration periods and runs failing data-quality criteria. Four major data sets—SET1 through SET4—have been analyzed, corresponding to exposures of approximately 60~days, 1.7~years, 3~years, and 6~years, respectively as indicated in Fig.~\ref{ref:expo}.

The analysis framework builds directly on the detector design described in the previous section. The LS veto plays an essential role in reducing backgrounds: radiogenic events from PMTs, nearby detector materials, or internal crystal contaminants are efficiently tagged by requiring no coincident signal in the surrounding LS or in neighboring crystals. The large LAB-based LS volume provides high-efficiency rejection of external gamma rays and multiple-scattering events, significantly improving the purity of the low-energy event sample used in WIMP searches.

After applying event-selection criteria, LS and inter-crystal vetoes, and calibration procedures, the resulting data sets form the basis for the model-dependent dark-matter analyses presented here.

The COSINE-100 WIMP search is primarily limited by low-energy backgrounds, especially PMT-induced noise that mimics scintillation signals in NaI(Tl). A dedicated noise-reduction pipeline is employed to suppress these instrumental backgrounds. This includes pulse-shape–based rejection of fast noise, asymmetry checks using the dual-PMT readout of each crystal, and removal of events with anomalous temporal or spatial characteristics.

For the model-dependent WIMP search, the remaining background components are further constrained using multivariate classification methods. Discrimination based on event pulse shapes optimized for NaI(Tl) scintillation, boosted decision trees (BDTs), and neural-network classifiers are trained to improve separation among nuclear-recoil–like events, electron-recoil events, and noise background events near threshold. These multivariate selections provide the primary event sample for the COSINE-100 model-dependent analyses.

A precise understanding of the detector’s background is crucial. Background contributions from internal contaminants, cosmogenic isotopes, PMT components, and external radiation are modeled using a full \textsc{Geant4}-based simulation framework~\cite{COSINE-100:2024ola,Choi:2024ziz}. The simulations are tuned and validated with calibration data, yielding agreement with measured spectra at the level of a few percent. This precision enables reliable spectral fitting, signal extraction, and systematic-uncertainty control in the WIMP analysis.

Model-dependent WIMP extraction began with the SET1 (60-day) data set. Using a 2\,keV analysis threshold, COSINE-100 excluded the DAMA/LIBRA-phase-1 $3\sigma$ signal region for spin-independent WIMPs in the Standard Halo Model at 90\%\,C.L.~\cite{Adhikari:2018ljm}. The subsequent SET2 analysis (1.7 years) improved event selection and reduced the threshold to 1\,keV. Combined with a more precise background model, this led to substantially enhanced sensitivity. No signal consistent with dark-matter interactions was observed, and COSINE-100 excluded DAMA’s allowed regions under several interaction hypotheses within the Standard Halo Model~\cite{COSINE-100:2021xqn}.

For the SET3 analysis (3 years), COSINE-100 achieved a 0.7\,keV threshold, extending sensitivity to low-mass WIMPs~\cite{COSINE-100:2024wji}. In the signal region, the observed background consists of comparable contributions from internal crystal radioactivity (originating from crystal growth processes), cosmogenic activation prior to underground installation, and surface-related backgrounds associated with detector encapsulation. Improved noise rejection and refined background modeling led to an order-of-magnitude improvement in WIMP limits relative to earlier COSINE-100 results.

\begin{figure}[!htb]
  \begin{center}
      \includegraphics[width=0.45\textwidth]{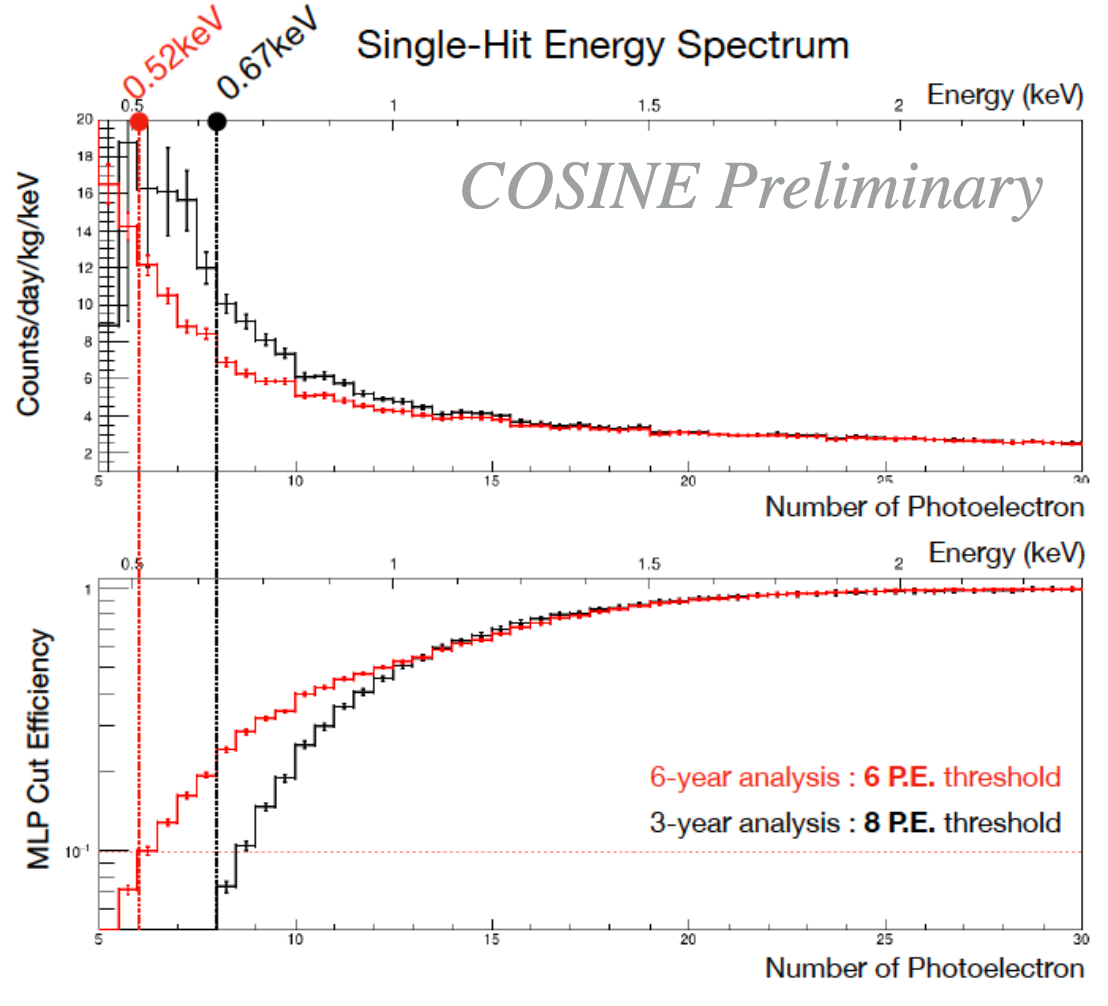}
  \end{center}
  \caption{Energy threshold improvement in the SET4 analysis.
    The SET4 analysis reduces the energy threshold down to 0.52 keV,
    nearly doubling the effective low-energy exposure.
    The figure shows the energy spectrum along with the selection efficiency,
    as a function of the photoelectron-equivalent energy.
  }
  \label{ref:fulldata}
\end{figure}

Analysis of the full SET4 (6-year) data set is in progress. Significant improvements in waveform classification—using simulated pulses for training and optimized multivariate variables—have enabled the analysis threshold to be lowered to 0.52\,keV as showen in Fig.~\ref{ref:fulldata}. This development nearly doubles the effective exposure available for the WIMP search. Final SET4 results are expected in early 2026.

\begin{figure}[!htb]
  \begin{center}
      \includegraphics[width=0.45\textwidth]{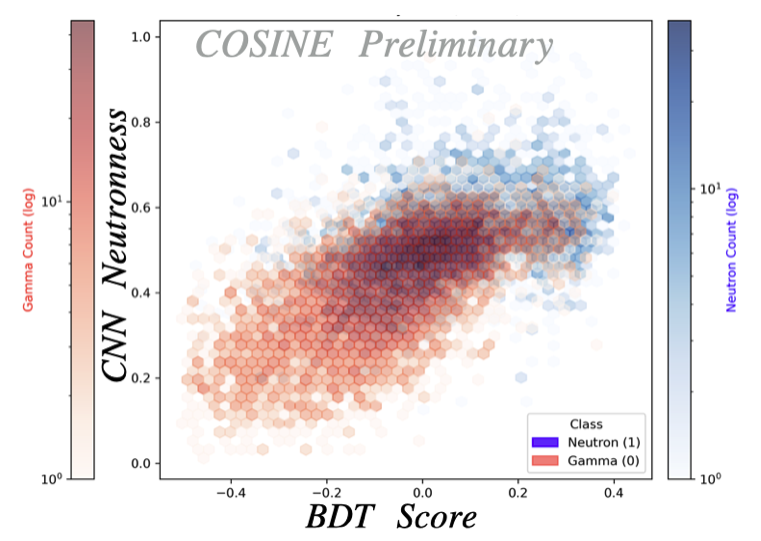}
  \end{center}
  \caption{Pulse-shape discrimination in COSINE-100.
The plot shows the two-dimensional PSD parameter space for neutron-induced (signal-like) and gamma-induced (background-like) events. The vertical axis corresponds to the CNN-based PSD output, while the horizontal axis shows the BDT-based PSD output. Events in the lower-left region are more gamma-like, whereas those in the upper-right region are more neutron-like.
  }
  \label{ref:psd}
\end{figure}

In parallel with the model-dependent searches, COSINE-100 is developing a dedicated WIMP-focused pulse shape discrimination (PSD) analysis to improve sensitivity to nuclear recoils in NaI(Tl). The method uses neutron-scattering data to model nuclear-recoil pulse shapes and gamma-induced events for electron recoils~\cite{Kim:2016igj,Joo:2018hom}. Preliminary studies in Fig.~\ref{ref:psd} show that Convolutional Neural Network (CNN)-based and BDT-based classifiers can separate the two populations better, indicating strong potential for future improvements when integrated with the standard analysis pipeline.

COSINE-100 has also examined alternative dark-matter interaction hypotheses, including isospin-violating scenarios and a range of effective field theory operators, using the same 1\,keV threshold and adopting updated NaI(Tl) quenching factors used by DAMA/LIBRA~\cite{COSINE-100:2021xqn}. In all cases, the COSINE-100 data show no excess above the expected backgrounds that could reproduce the DAMA annual modulation signal within the Standard Halo Model. These null results rule out such interpretations within the tested interaction frameworks.

In addition, COSINE-100 has initiated searches for Migdal-effect signals, which provide sensitivity to sub-GeV dark matter. For low-mass WIMPs, nuclear recoil energies fall below threshold, but the accompanying Migdal electron or gamma emission produces a signal detectable in NaI(Tl). This extends the experiment's sensitivity beyond the traditional nuclear recoil channel~\cite{COSINE-100:2021poy}.

Finally, COSINE-100 enables a broad range of searches for physics beyond the Standard Model, including constraints on bosonic super-WIMPs~\cite{COSINE-100:2023dir}, boosted dark matter~\cite{COSINE-100:2023tcq,COSINE-100:2018ged}, inelastic dark matter scattering on $^{127}$I~\cite{COSINE-100:2023jts}, and possible annual modulation signatures from solar bosonic dark matter~\cite{COSINE-100:2023uku}. These studies benefit from the detector’s low background rate, large exposure, and well-characterized response, enabling competitive limits across a broad range of dark-sector models.

\section{The COSINE-100 Upgrade}

The light yield of NaI(Tl) scintillation crystals increases substantially as the operating temperature decreases. Cooling the crystals to approximately $-35^\circ$C enhances photon production, leading to improved energy resolution and a lower analysis threshold~\cite{Lee:2021aoi}. This strong temperature dependence motivates further optimization of low-temperature detector operation for future phases.

\begin{figure}[!htb]
\begin{center}
\includegraphics[width=0.4\textwidth]{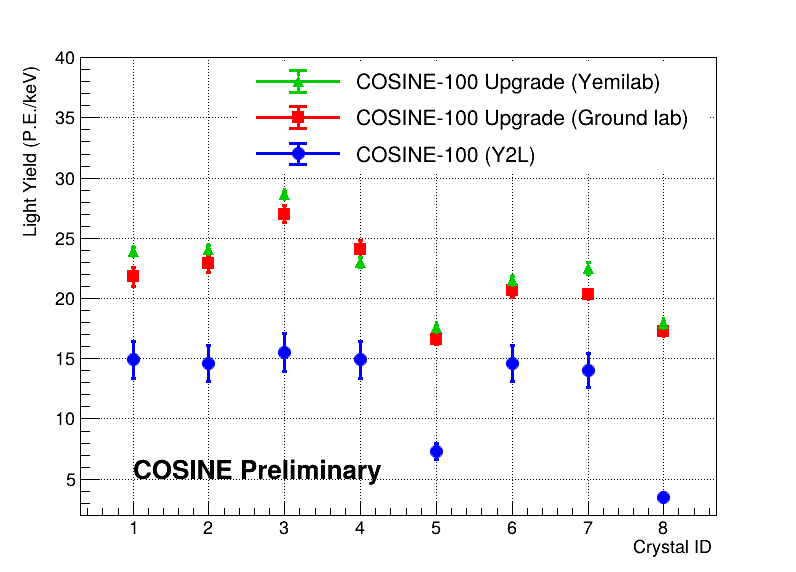}
\includegraphics[width=0.4\textwidth]{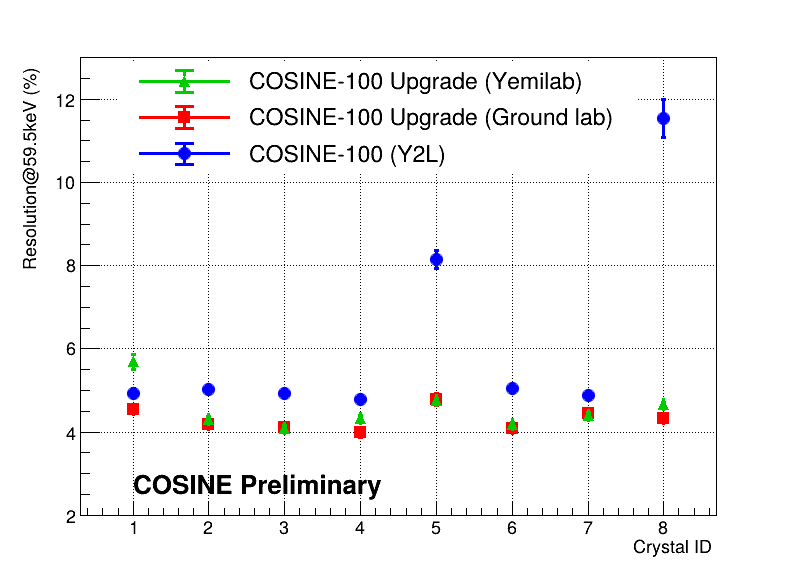}
\end{center}
\caption{COSINE-100 Upgrade crystal detectors.
The upper panel compares the light yields of eight crystals measured in three configurations: the original COSINE-100 at Y2L, the COSINE-100 Upgrade in the ground laboratory, and the COSINE-100 Upgrade at Yemilab. More than a 50\% increase in light yield is observed after the upgrade. The lower panel shows the energy resolution of the 59.5~keV $^{241}$Am gamma peak, which also demonstrates significant improvement in the upgraded detectors.}
\label{ref:uprnd}
\end{figure}

The COSINE-100 experiment has recently been relocated to its new home at Yemilab, where the upgraded detector incorporates in-house crystal machining, polishing, and re-encapsulation. These improvements significantly enhance optical coupling between the NaI(Tl) crystals and the PMT photocathodes. Figure~\ref{ref:uprnd} shows the re-encapsulated crystals exhibit approximately a 50\% increase in light yield and about a 10\% improvement in energy resolution~\cite{Choi:2020qcj}. To optimize these performance gains, the total crystal mass was slightly reduced from 106 kg to 99 kg, and updated encapsulation designs were introduced to ensure stable, low-background operation~\cite{Lee:2024wzd}.

\begin{figure*}[!htb]
\begin{center}
\includegraphics[width=0.95\textwidth]{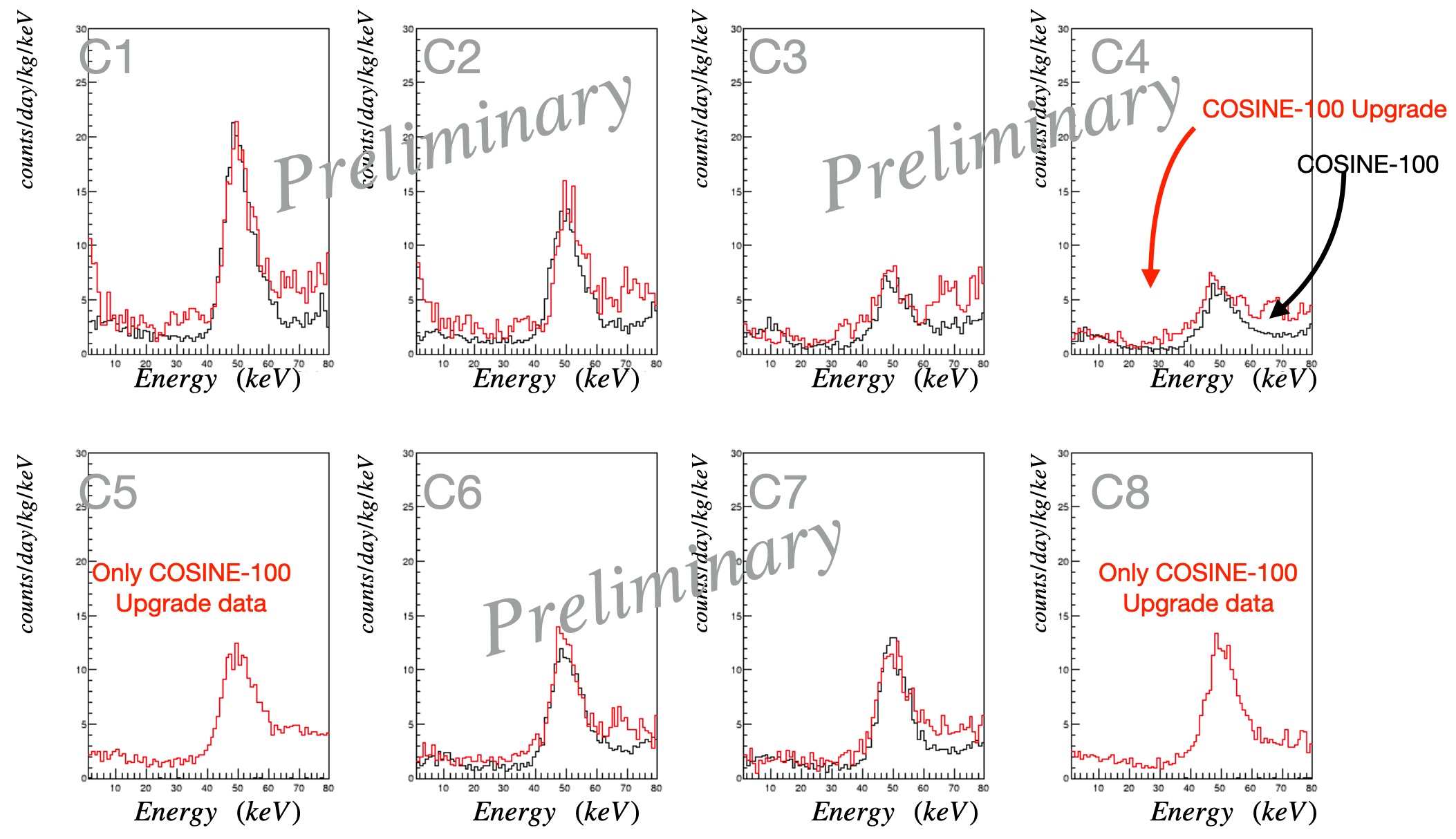}
\end{center}
\caption{Initial test data from the COSINE-100 Upgrade.
  Energy spectra from each upgraded crystal are compared with those from the original COSINE-100. Below 10 keV, the performance is comparable to the previous detector, and crystals 5 and 8—which were not operational in COSINE-100—have been successfully recovered.
}
\label{ref:initdata}
\end{figure*}

Extensive tests of new encapsulation methods have been performed to ensure mechanical robustness, optimize light transmission, and minimize material-induced backgrounds. These developments were implemented in the COSINE-100 Upgrade now being tested at Yemilab. Installation has been finished with new liquid scintillator for veto and background rejection systems~\cite{Kim:2024spf}. Early data from the upgraded detector show background levels comparable to those of the original COSINE-100, reaching approximately 3~counts/day/kg/keV below 10~keV. Notably, two large modules (crystal 5 and crystal 8) that were unusable in COSINE-100 have now been successfully recovered and integrated into the new array as shown in Fig.~\ref{ref:initdata}.

\begin{figure}[!htb]
\begin{center}
\includegraphics[width=0.45\textwidth]{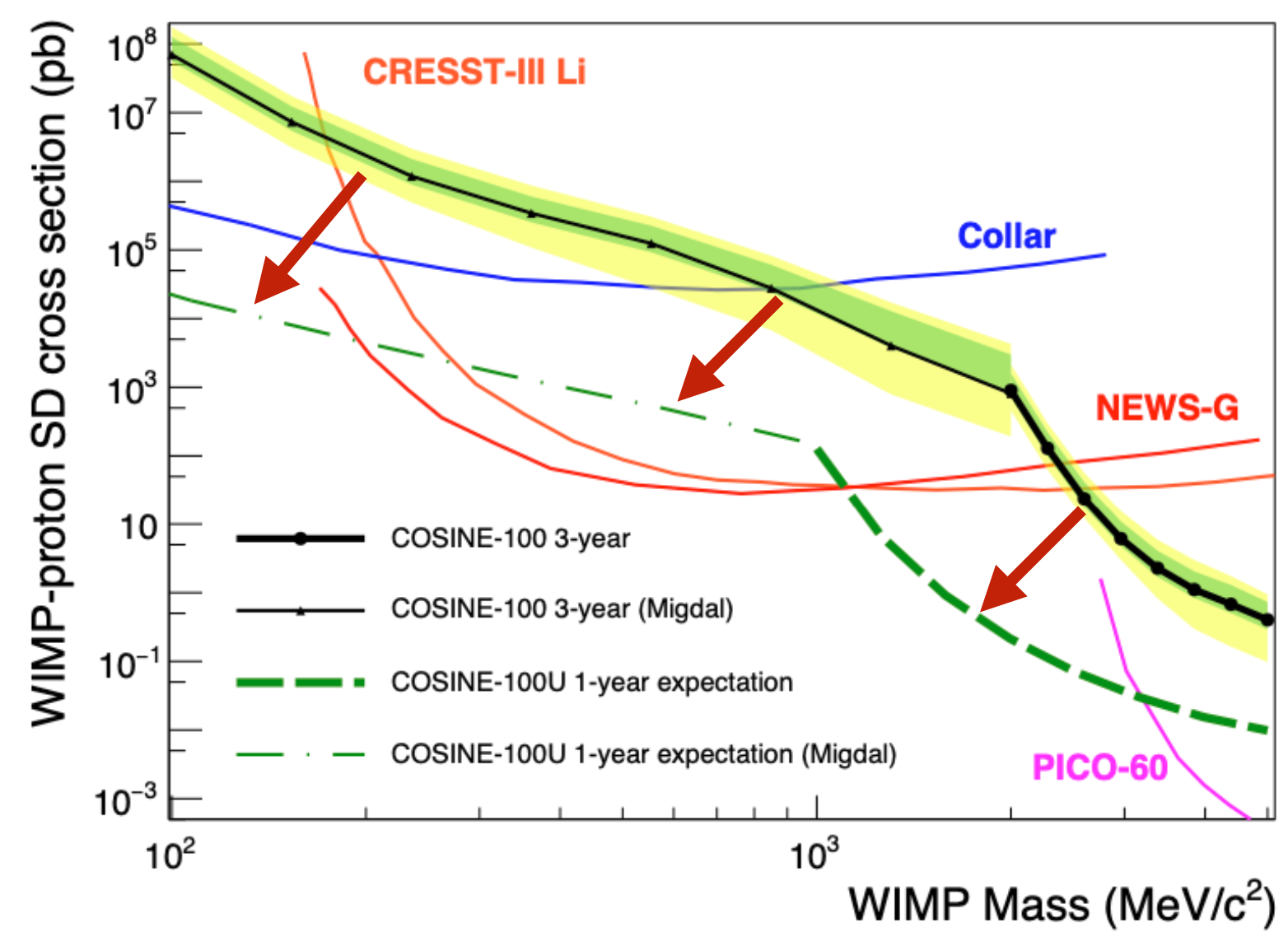}
\end{center}
\caption{Expected sensitivity of the COSINE-100 Upgrade.
  Compared to the original COSINE-100 results, the upgraded detector is projected to significantly improve sensitivity to low-mass spin-dependent WIMP interactions, as indicated by the green dotted curves.
}
\label{ref:expsens}
\end{figure}

With these enhancements, the COSINE-100 Upgrade is expected to deliver significantly improved sensitivity to spin-dependent WIMP interactions as shown in Fig.~\ref{ref:expsens}. The higher light yield, improved crystal–PMT coupling, and better-performing detector modules collectively strengthen the search reach.

\section{ Summary and Outlook}
The COSINE experiment seeks to directly detect dark matter through WIMP-induced nuclear recoils using NaI(Tl) scintillation crystals, enabling a model-independent test of the DAMA modulation claim. COSINE-100 collected 360 kg$\cdot$year of data from 2016 to 2023, operating stably for 6.39 years with over 91\% good-quality runs.
Analyses using up to 6 years and thresholds of 0.5–0.7 keV are underway to test a DAMA-like modulation within the Standard Halo Model. The COSINE-100 Upgrade, now installed at Yemilab, incorporates in-house crystal machining, polishing, and re-encapsulation, improving light yield by about 50\% and energy resolution by roughly 10\%. Operating at $-35,^\circ\rm C$ further enhances scintillation output. Early test data exhibit similar background levels as COSINE-100, with previously unusable detectors successfully recovered. These improvements are expected to strengthen sensitivity to spin-dependent WIMP interactions.

Looking forward, COSINE-200 is being developed with a larger crystal mass and a target background level below 1~counts/day/kg/keV~\cite{Lee:2023jbe}. The ongoing in-house production of large, high-purity NaI(Tl) crystals will enable scalable, low-background detector fabrication. These advancements position COSINE to achieve competitive sensitivity to low-mass and spin-dependent WIMP interactions and to play a leading role in the next generation of direct dark matter searches.

\section{Acknowledgments}
We thank the Korea Hydro and Nuclear Power (KHNP) Company for providing underground laboratory space at Yangyang.

\section*{FUNDING}
This work is supported by the project code NRF2021R1A2C1013761, Republic of Korea.

\section*{CONFLICT OF INTEREST}
The author declares that he has no conflicts of interest.

\nocite{*}

%

\end{document}